# DVB-S2x Enabled Precoding for High Throughput Satellite Systems

**Pantelis-Daniel Arapoglou, Alberto Ginesi, Stefano Cioni, ESA/ESTEC**

**Stefan Erl, Federico Clazzer, DLR**

**Stefano Andrenacci, Alessandro Vanelli-Coralli, University of Bologna**

## Abstract

Multi-user Multiple-Input Multiple-Output (MU-MIMO) has allowed recent releases of terrestrial LTE standards to achieve significant improvements in terms of offered system capacity. The publications of the DVB-S2x standard and particularly of its novel superframe structure is a key enabler for applying similar interference management techniques -such as precoding- to multibeam High Throughput Satellite (HTS) systems. This paper presents results resulting from European Space Agency (ESA) funded R&D activities concerning the practical issues that arise when precoding is applied over an aggressive frequency re-use HTS network. In addressing these issues, the paper also proposes pragmatic solutions that have been developed in order to overcome these limitations. Through the application of a comprehensive system simulator, it is demonstrated that important capacity gains (beyond 40%) are to be expected from applying precoding even after introducing a number of significant practical impairments.

## I. Introduction – The Need for Precoding

A variety of single-user and multi-user Multiple-Input Multiple-Output (SU/MU-MIMO) transmission modes (TMs) has enabled consecutive releases of Long Term Evolution (LTE) and LTE-Advanced to achieve quantum leaps in terms of offered spectral efficiencies [1], [2]. Although in early releases of LTE only elementary MU-MIMO TMs are present, Release 10 onwards fully developed (downlink) MU-MIMO for up to 8 antennas because:
- Most of the user terminals (UTs) are not able to support a large number of receive antennas due to terminal size/complexity.
- MU channels are not susceptible to high spatial antenna correlation as the corresponding SU ones.

The satellite community has witnessed a similar capacity push for its interactive satellite networks leading to a proliferation of High Throughput Satellite (HTS) systems. The total system throughput achieved by such systems is in the order of hundreds Gigabits/s with future (2025) generations of large satellites targeting up to one Terabit/s [3]. HTS systems typically employ the Ka frequency band (20/30 GHz) on the link from the satellite towards the UTs forming multiple spot beams on ground and re-using the available system bandwidth based on a certain pattern in frequency and polarization (colouring scheme, see Figure 1). Current large HTS systems employing cutting edge technology –like Viasat-1 with a throughput of about 140 Gbit/s– typically split the available bandwidth in two frequency bands and two orthogonal polarizations generating the so-called *four colour* beam pattern across the coverage area. The corresponding spectrum necessary to support the user services needs to be available to the feeder link, that is the link between the gateway stations (GWs) and the satellite.

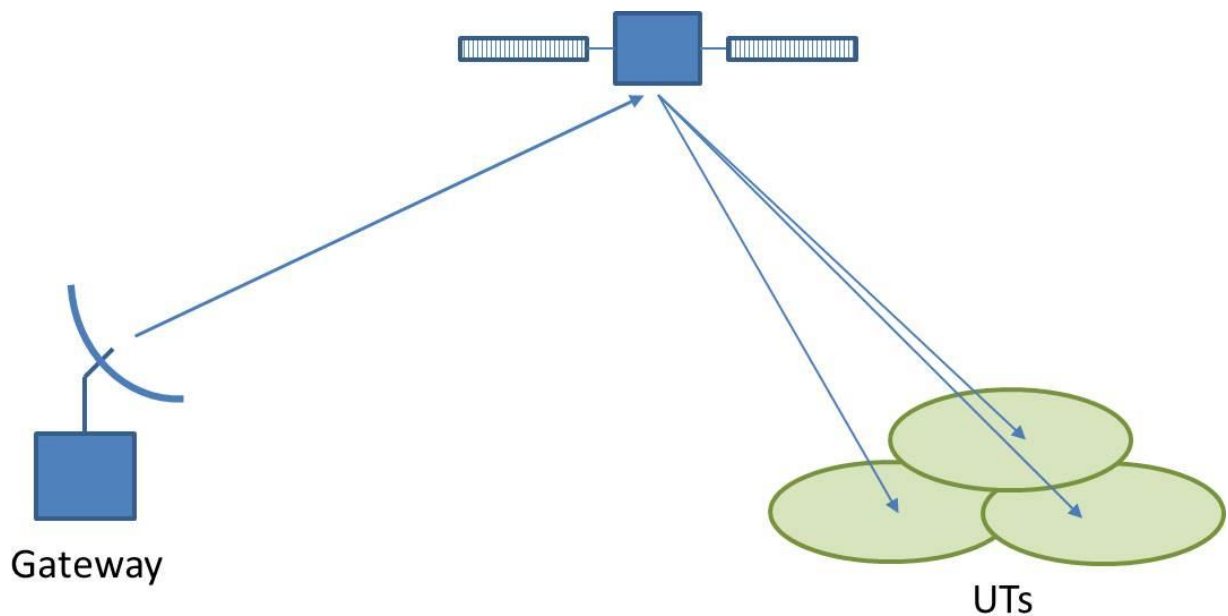

Figure 1 Multibeam HTS system architecture.

In terms of waveform, due to its high number of modulation and coding schemes (ModCods), its available mechanisms to adapt the ModCods to the channel conditions (ACM) and its high performance, Digital Video Broadcasting via Satellite 2nd Generation (DVB-S2) [4] has been by far the most popular choice for providing fixed satellite services (FSS) through HTS systems. DVB-S2 is based on amplitude and phase shift keying (APSK) modulation and BCH/LDPC (Bose Chaudhuri Hocquenghem / Low Density Parity Check) concatenation of channel codes.

Looking at the limited spectrum available in Ka-band for FSS UTs, further leaps in terms of HTS system capacity seem only possible by diverging from the paradigm of the four colour frequency re-use and moving instead to a new paradigm of higher frequency re-use (of 2 or even full frequency re-use of 1). Under such an arrangement of the frequency assignments, the interference environment between the co-channel beams becomes harsh, as there is no longer spatial isolation between co-channel beams and their side lobes. This negates any potential capacity increase stemming from the use of the additional spectrum as the high co-channel interference leads to very low signal-to-noise plus interference ratio (SNIR) and, in turn, to low spectral efficiency.

Following in the footsteps of terrestrial LTE, MU-MIMO seems like an excellent solution to manage this high intra-system interference originating from the co-channel beams in the system. Indeed, MU-MIMO techniques have been proposed for the forward link of HTS broadband interactive systems and are collectively referred to as *precoding* [5]. That is, the GW on ground precodes across the signals intended to the UTs that are distributed over a high number of spot beams via a multi-feed satellite antenna.

Following the early works of [6], [7], [8], the literature on satellite based precoding has intensified recently [9], [10] demonstrating very high theoretical precoding gains in terms of system capacity even when sub-optimal linear precoding techniques are applied over the multibeam fixed satellite channel. Despite these promising results, only very recently researchers have started looking into the challenges of implementing precoding in a DVB-S2 based HTS practical system with an aggressive frequency re-use [11]. These implementation challenges, which are unfolded in Section II, are the main focus of this paper which reports the work carried out in the last phase of the ESA Artes 1 activity "Next Generation Waveform for Improved Spectral Efficiency (NGW)" [12]. Many of the difficulties have been accommodated by the novel superframe format that was introduced in the extension of DVB-S2 (DVB-S2x) [13]. How the superframe has helped in overcoming issues related to precoding is explained in Section III. The rest of the fundamental implementation challenges for applying precoding over HTS systems are related to multicast (or frame-based) precoding and to the UTs synchronization strategy, topics which are addressed in Section IV and Section V, respectively. After modeling these features, the performance of a practical system making use of the multicast precoding technique is compared against a reference four colour system in Section VI. Useful conclusions are drawn in Section VII.

## II. Implementation Challenges in Practical Systems

The support of classical (LTE-like) precoding algorithms in DVB-S2-based networks presents a number of challenges ultimately due to the fact that the DVB-S2 standard had been conceived with broadcast Direct-to-Home (DTH) services as the primary application. In the following paragraphs, a number of such implementation challenges are discussed individually.

### Multicast PLFRAME

In the structure of Figure 2 corresponding to the DVB-S2 physical layer frame (PLFRAME), the Layer 2 packets of different users are multiplexed together within a single codeword. In addition, their bits are interleaved together by the DVB-S2 bit interleaver so that each resulting channel symbol might stem from the mapping of bits belonging to different users. As precoding works at symbol level (and thus needs to match a given channel symbol to a given user) it would be virtually impossible to generate a classical precoding matrix this way. In addition, within the PLFRAME there are bits which do not belong to any specific user, as they are meant to be multicasted to all the users of the PLFRAME: these are the PLH (physical layer header) bits as well as the LDPC and BCH parity redundancy bits.

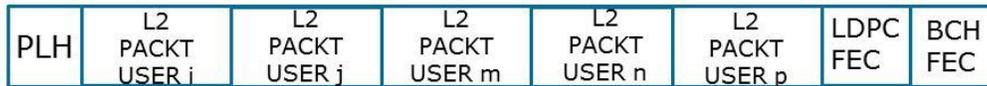

Figure 2 Structure of a DVB-S2 PLFRAME.

From a precoding point of view, this approach to DVB-S2 PL framing is adding an important practical constraint since it implies that the precoder cannot be designed on a user-by-user basis (conventional unicast precoding). Rather, some type of `equivalent' frame based precoding should operate on the channels of multiple UTs encapsulated in a frame. This has given rise to a new research area within the satellite community called *multicast precoding* [14], which involves a user selection and a non-conventional precoding method. Multicast or frame based precoding approaches are briefly discussed in Section IV.

## PLFRAME of Variable Length

The DVB-S2 PLFRAME size at the output of the encoder is constant and equal to either 16k or 64k bits [4]. Nevertheless, depending on the selected modulation, the size of the resulting PLFRAME in symbols is variable. In a multibeam HTS system where ACM is employed, it turns out that every beam transmits a different frame size in symbols. To explain how this comes about, let us consider Figure 3.

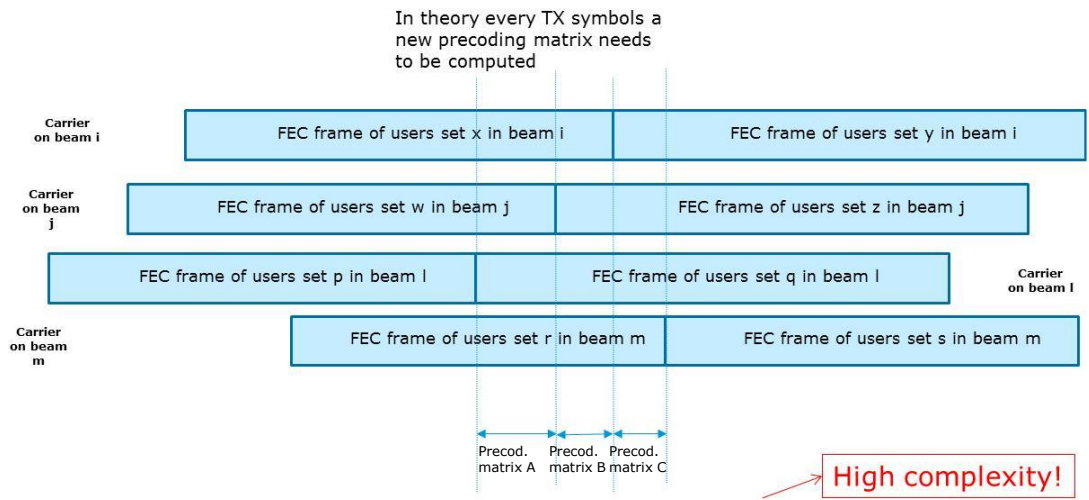

Figure 3 Mis-alignment of the DVB-S2 PLFRAMES across co-channel carriers when using ACM mode.

In this figure, the PLFRAMEs related to different DVB-S2 carriers serving different co-frequency beams are depicted. In this configuration, due to the variable PLFRAME length when using ACM, even if at system initialization the PLFRAMEs across the 4 carriers were aligned in time, the alignment would soon be violated. As a consequence, the rate of the precoding matrix computation could approach the symbol rate (as illustrated in the same figure). Also the pilot symbols necessary to estimate the channel would be misaligned. This issue calls for a more regular physical layer framing structure. A solution along this line has been implemented in the Annex E of DVB-S2x [13], through the exploitation of the Bundled PLFRAMEs. This is further elaborated in Section III of this paper.

## Imperfect Channel Estimation

Precoding assumes the knowledge of the UTs channel state information (CSI) at the GW transmitter. CSI should be available at the GW so that multiuser precoding can be performed. However, in the case of precoding, each UT needs to estimate a whole vector of channels –the so called channel direction information (CDI)- instead of a single element in the channel matrix. In addition, as already the case in existing DVB-S2, each UT should provide to the GW also the channel quality indicator (CQI), i.e. its SNIR. These channel estimation operations take place under a higher interference environment (compared to the usual four colour systems) due to the more aggressive interference reuse, with the strongest interference approaching the same power as the main carrier and a number of additional non-negligible interferers. A possible distribution of C/I and C/N for a given geographical point over a full-frequency re-use network, is represented in Figure 4.

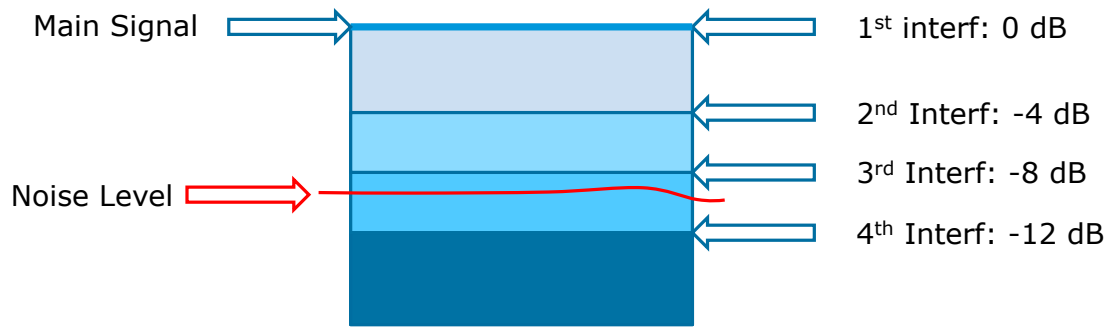

**Figure 4 A possible distribution of C/I and C/N for a geographical location in a full frequency re-use network.**

In this case, the receiver needs to collect CSI on every single carrier with not negligible power. This implies locking onto and performing channel estimation on the main signal, the first interference at 0dB, the second interference at -4dB, the third interference at -8dB, and finally the fourth interference at -12 dB. In other words, the receiver would need to perform frame, carrier, and timing synchronization as well as channel amplitude and phase estimation on a signal which is 12 dB below the carrier and possibly a few dBs below noise, which is a non-trivial task.

This calls for a much more sophisticated UT synchronization and channel estimation approach that has been developed during the ESA NGW R&D activity [12] and is overviewed in Section V. Despite putting in place these mechanisms, the CSI estimate still contains residuals errors which reduce the expected gains of the technique. Along similar lines, also the number of subchannel coefficients that is included in the CDI is limited, due to the limited capability of estimating below a certain carrier-to-interference level. The effect of both imperfect CSI estimation as well as the threshold effect below which coefficients cannot be accessed on the system performance is reported in the system simulation results presented in Section VI.

### Outdated Phase Estimate

Apart from the quality of the CSI estimate, a major physical constraint in GEO based satellite systems is the latency involved in the feedback due to the long round trip time, which amounts to approximately 500 ms. During this time, a number of sources could potentially change the phase estimate compared with what was estimated at the UT. The reason only the phase estimate variation is of concern is the slow varying nature of the fixed satellite channel amplitude when impaired by atmospheric effects. On the contrary, there is a plethora of factors with relative fast phase dynamics including the phase noise of the local oscillators (LOs) on board and on ground (particularly the UT LNB), the change of slant path geometry due to the satellite movement or due to the spatial separation between the satellite antenna feeds.

After studying all these effects in the frame of the NGW activity, it turns out that the most relevant phase variations are the ones between the on board payload chains forming the multiple spot beams on the Earth's surface. The effect of the phase variation due to the payload chains are due to the different LOs in charge of the uplink to downlink frequency conversion. Nevertheless, the dynamics of the phase changes can be controlled by using a configuration where a single stable oscillator provides a common reference to the individual frequency converters; this is not an uncommon multibeam satellite payload architecture. On the other hand, the phase variation due to the LNB is in common for all the channel matrix coefficients measured by a given UT. It is then easily demonstrated that the contributions due to receiver LNB phase variation do not impact the received SNIR and that this contribution can be omitted.

### Multiple Gateways

Typically, the literature on satellite-based precoding assumes that all spot beams in the system are served by a single GW, which is not feasible due to the limited feeder link spectrum. This is the case even in a conventional four colour system when a large number of beams is used. The situation in terms of high number of GW stations is aggravated by moving to more aggressive frequency re-use architectures, although there are efforts to move the feeder link to higher frequency bands [15] to moderate the number of GWs.

The effects of multiple GWs on the precoding design lead to a performance loss with respect to single GW precoding due to the fact that there is no longer a single transmitter entity in possession of the CSI for all UTs. This allows the GW to pre-compensate for the co-channel interference of the UTs only for the subset of beams that this GW serves. This fact, the impact of which is quantified in Section VI, along with increase of the on board equipment when higher frequency re-use factors are considered, renders precoding much more suitable for regional HTS systems providing multibeam coverage through a moderate number of spot beams.

### Impact on availability

Precoding strives to increase system capacity by enhancing UTs with good channel conditions and penalizing UTs that are experiencing unfavourable channel conditions. Obviously, this will reduce the availability of the latter UTs. Moreover, due to the system moving from a conventional frequency re-use of four to a frequency re-use of one, a deterioration of the carrier-to-interference ratio (C/I) is inevitable, thus resulting in low SNIR levels that may not be supported by DVB-S2 ModCod options. Fortunately, DVB-

S2x has addressed this issue by extending the DVB-S2 ModCods to lower SNIR ranges with the help of BPSK and a spreading operation. This extension is expected to allow systems with precoding to preserve the availabilities typically required by system operators down to -10dB of SNIR.

## Payload mass implications

Precoding requires frequency re-use 2 or full frequency re-use networks. The implication to the satellite payload in terms of number of High Power Amplifiers (HPA) might be significant for conventional Travelling Wave Tube Amplifier -based multi-beam systems. Indeed, the duplications or quadruplications of transmitter chains might represent a formidable limitation to the successful deployment of HTS pre-coded networks with passive antennas. Instead, when using an architecture with active or semi-active antennas with one HPA per feed, the payload mass is not impacted by the choice of the number of colours in the network and thus precoding can be easily deployed. These architectures will also become more appealing as the power and efficiency of Ka-band Solid State Power Amplifiers (SSPA) improves. Other scenarios with passive antennas where precoding could be applied is regional class of HTS networks, where the satellite platform might be selected due to power and antenna accommodation requirements, and thus the system might not be strictly mass-limited.

## III. Summary of the DVB-S2x Superframe Option

A novel superframing structure is described in Annex E of the DVB-S2x standard [13]. This is thoroughly discussed in another paper within this same special issue. The objective of this section is to relate the characteristics of the superframe of Annex E to the issues identified in the previous section. To start with, we consider Figure 5, where the format with bundled PLFRAMES of DVB-S2x Superframe structure is illustrated.

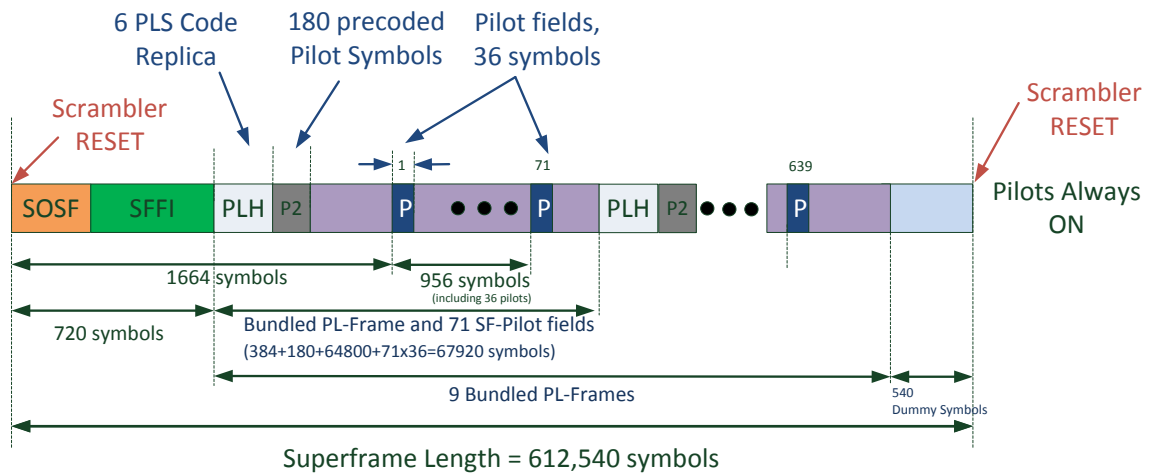

Figure 5 Super-frames of format with bundled PLFRAMEs (64 800 payload size) [13].

The first feature to note is the relatively long SOSF field (Start of SuperFrame- 270 symbols) together with pilots sequences (SF-Pilots, indicated as P in Figure 5) interspersed along the payload of the Superframe container. These fields are also conveniently filled by pre-defined sequences of symbols derived by a set of 256 and 32 Walsh-Hadamard sequences, respectively. These sequences have the property of being quasi-orthogonal meaning that if suitably distributed across the co-channel carriers, they consent to support efficient synchronization and channel estimations even in the presence of very strong interference. Naturally, this is true as long as their respective superframes containers are kept time-aligned. A detailed discussion of the feasibility of such alignment has been elaborated in the previous section. Here it is worth to note that thanks to the constant properties of the superframe, the alignment can be virtually kept constant throughout the service (if one neglects the very small time variation of payload differential group delay across different transponders).

For the sake of the synchronization and estimation techniques that will be discussed in Section V, it is fundamental to note also that

- SOSF and SF-Pilots are not-precoded;
- SOSF and SF-Pilots consist of beam-specific orthogonal Walsh-Hadamard sequences;
- two different scrambling sequences are applied to the superframe: the first sequence, the so called *reference data scrambler*, is the same for all beams, it overlays only the SoSF and SF-Pilots, and is restarted at each Start of Super Frame. The second scrambling sequence, the so called *payload data scrambler*, is beam dependent, overlays the data payload, and provides resilience to co-channel interference.
- the superposition of the beam-specific Walsh-Hadamard sequences in the SOSF and SF-Pilots and the *common Reference Data Scrambler* yields a unique beam-specific *signature* that can be used for waveform/beam identification.

Thanks to the exploitation of the Bundled PLFRAMEs, the time alignment of superframes among different co-frequency carriers also allows to keep the payload data aligned. The possible structures of the Bundled PLFRAMES are illustrated in Figure 6. The concept is that by carefully arranging a number of PLFRAMES together according to the modulation format, a constant length bundle is achieved. If a multicast precoding coefficient is computed for the entire bundle, effectively the rate of computing the multicast precoding matrix is equal to the Bundled PLFRAME rate. Therefore, the complexity of the precoding matrix computation is minimized. However, the drawback implied by the use of the relatively long Bundled PLFRAMES is represented by possible inefficiencies of the scheduler as well as potential increase of data delay jitter.

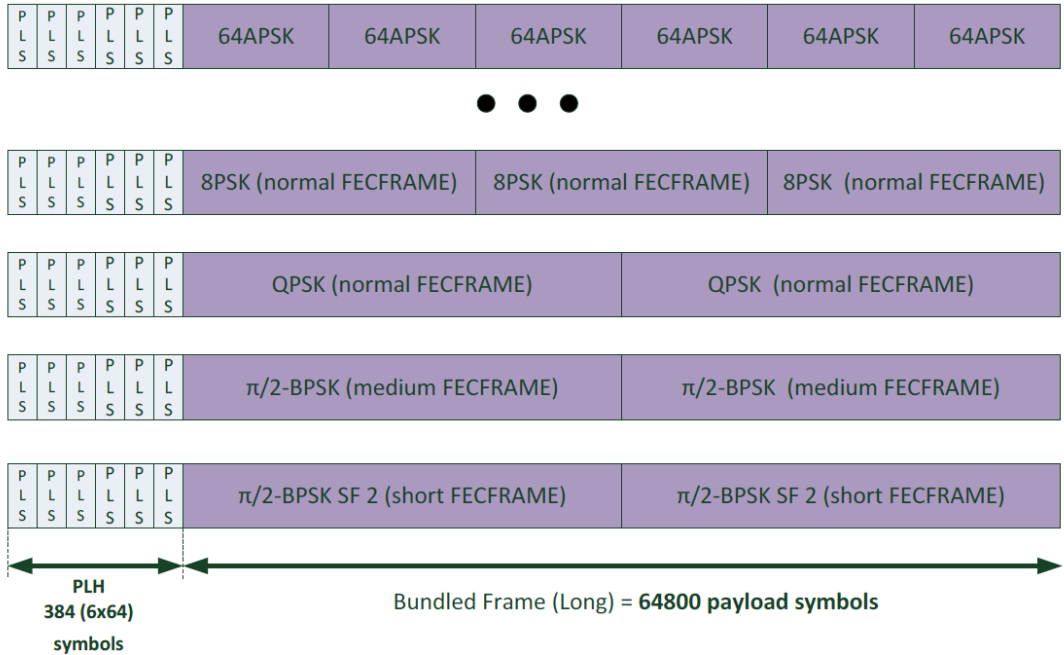

Figure 6 Selected Examples of Bundled PLFRAMEs (64 800 payload size, pilots not shown).

Incidentally, Figure 5 shows other types of pilots that are present at the level of the Bundled PLFRAMES, i.e the P2 pilot sequences. As opposed to the former pilots sequences and SoSF, the latter are precoded with the same precoding matrix used over the data symbols and, as it will be explained in Section V of the paper, they are used to assist data detection of the main carrier. In order to properly support channel amplitude and phase detection these pilots carry the same modulation format of the payload data (this feature is desirable when in the presence of non-linear distortions). The different constellation points of the considered modulation are conveniently represented within the sequence with a fair share of time.

## IV. Multicast Precoding Approaches

To address the multicast PLFRAME issue, a new family of precoding algorithms have been recently developed referred to as *multicast* or *frame based precoding* algorithms [14]. The term implies that the same precoding matrix is applied over the symbols of a bundled PLFRAME corresponding to multiple UT channels in each spot beam. This approach to precoding is fundamentally different compared to the application of MU-MIMO in LTE and LTE-Advanced.

The aim of this section is not to enter into the detail of the new family of multicast precoding techniques, but provide instead references to the corresponding work and support the following sections. Three new multicast precoding techniques, an overview of which is presented in [14], [16] address both the issue of how to select the UTs that are grouped in the same frame as well as how to calculate the precoding matrix. The three approaches differ in both these respects and consist of (in increasing degree of complexity):

- Geographical user selection and MMSE (minimum mean square error) precoding [17].
- A variant of the block singular value decomposition presented in [18].
- A multicast aware user scheduling along with a weighted max-min fair optimization for obtaining the precoding matrix [19], [20].

For the system level performance evaluations that are presented in Section VI, the first option of [17] has been implemented due to its simplicity and encouraging results that it offers. Nevertheless, it is worth mentioning that only linear precoding is employed in all three cases. The reason is that despite not reaching the capacity achieving Dirty Paper Coding [21], linear precoding grasps already most of the gain precoding has to offer, a conclusion that has been witnessed also in terrestrial wireless communication systems [22].

## V. Terminal Synchronisation Strategy

Before entering into the details of the synchronization strategy, it is worthwhile noting that the objective of designing the synchronization chain for a multicast precoded system is far from being a trivial application of already known procedures. On the one hand, multicast precoding is in fact designed to reduce, ideally remove, the interference generated by the co-channel beam waveforms, thus making auxiliary parameters estimation almost impossible on the precoded interferer waveforms. On the other hand, non-precoded SOSF and SF-pilots are subject to extreme high interference levels, much higher than that experienced in non-precoded systems. As a matter of fact, if that would not be the case, precoding would not be necessary in the system itself. Therefore, we can safely state that synchronization and estimation in precoded HTS systems is a formidable challenge by design.

The composite received signal, $y_k[n]$, at the k-th UT consists of the superposition of waveforms transmitted through B different interfering beams and it can be modeled as

$$y_k[n] = \left(\sum_{b=1}^{B} h_{kb}[n]x_b[n - \tau_{kb} - \tau_d n]e^{-i(2\pi\Delta f_{kb}nT + \theta_{kb})}\right)e^{-i(2\pi(f_o + \frac{1}{2}f_d n)nT + \varphi[n])} + n_k \qquad (1)$$

where $h_{kb}[n]$ is the complex channel coefficient at discrete sampling time n, $x_b[n]$ is the signal trasmitted through the b-th antenna beam, $\tau_{kb}$, $\Delta f_{kb}$, and $\theta_{kb}$ are the time, frequency, and phase offsets of the b-th received waveform at the k-th UT, respectively. Moreover, due to the characteristics of the receiver, the composite received signal is affected by the following common impairments which depend only on the considered UT: time and frequency drift, i.e., $\tau_{dk}$, $f_{dk}$, frequency offset $f_{ok}$, and phase noise $\varphi_k[n]$.

Let us consider an indexing function i(b) taking values in the set $S = \{0,1,2,\ldots,B\}$, i.e., $i(b) \in S = \{0,1,2,\ldots,B\}, b = 1,2,\ldots,B$, where $i(b) = 0$ means that the b-th waveform component is not considered in the processing. In order to be able to estimate the CSI to be sent back to the gateway, a UT shall proceed with the following general synchronization/estimation operations:
  a) identify the frame boundary, i.e., frame synchronization, for the i(b)-th waveform component, $b = 1,\ldots,B$;

b) estimate frequency, phase, and time for the i(b)-th waveform component, $b = 1, \ldots, B$;
c) perform channel estimation for the i(b)-th waveform component, $b = 1, \ldots, B$ to be sent back to the gateway.

To this aim, the k-th UT shall:
- perform and apply a coarse frequency estimation by means of a non-data aided estimator on $y_k[n]$, since the frame synchronization procedure is not performed yet;
- for each waveform $i(b)$, $b = 1,2,\ldots,B$, perform frame synchronization, if pilot fields or unique words are present in the frame format, to identify frame boundaries. Non-coherent post detection integration can be applied to cope with the residual frequency uncertainty and the time varying phase impairment; for those waveform components for which frame synchronization is not successfully achieved $i(b)$ is set to zero so as to exclude them from the subsequent processing;
- for each waveform $i(b)$, $n = 1,2,\ldots,B$ for which frame synchronization is successfully achieved, perform fine time tracking, phase and frequency tracking, and channel estimation.

A high level block diagram of the synchronization procedures described above is depicted in Figure 7.

It is worth mentioning that time tracking, phase and frequency tracking as well as CSI estimation procedure are performed using pilot assisted algorithms, which can take advantage of the orthogonal Walsh-Hadamard pilot sequences specified in Annex E of the DVB-S2x standard [13].

Finally, it is worthwhile noting that, assuming the same baud rate for all of the received waveforms, two possible implementations of the general procedure described above can be devised, depending on the relative delays of the received waveforms. One implementation shall address the case of low data rates for which the received waveforms can be considered "quasi synchronous", i.e., the relative time delays $\tau_{kb}$ are within one symbol time. The second implementation deals with the complementary case of "non-synchronous" waveforms, i.e., the relative time delays $\tau_{kb}$ may exceed one symbol time.

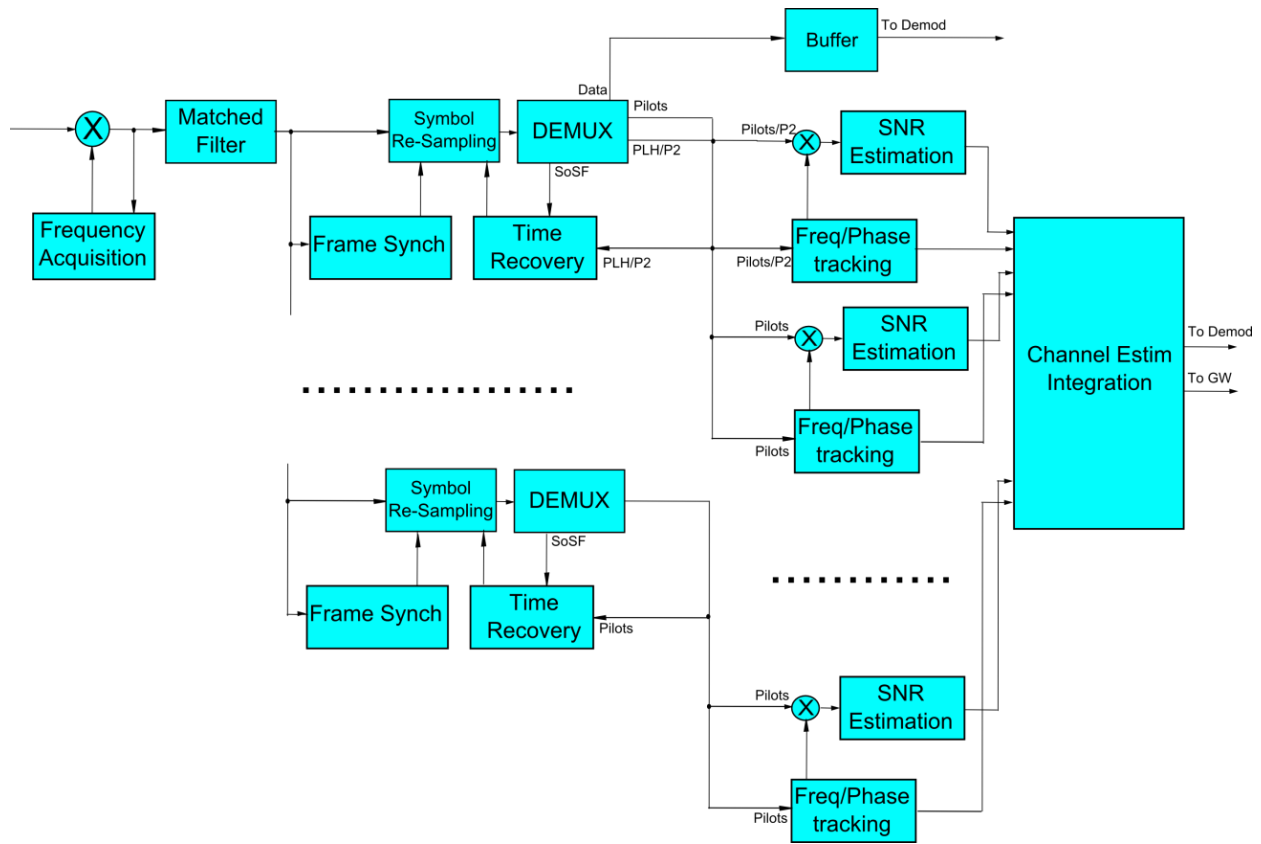

**Figure 7 Block diagram of synchronization procedure taking place at the UT in a system employing precoding.**

## VI. System level Performance Model & Results

The system level performance evaluation carried out in this Section aims at demonstrating the benefits of using precoding in a multibeam HTS system under realistic conditions compared to the classical 4 colours frequency reuse configuration. In this context, the precoding algorithm is applied not only under ideal conditions, that are used as a reference for the ultimate performance of the system, but also taking into account different sources of errors affecting the channel matrix. Specifically, the effect of imperfect channel estimates, outdated channel estimates and the effect of errors in the channel estimates are also considered (see also Section II).

Traffic scheduling is also taken into account in the simulation framework, i.e. how to efficiently schedule the transmission of users given that precoding is applied on a user cluster basis instead of a conventional user-per-user basis. The scheduling task is a complex problem that can become even more cumbersome when precoding enters the picture. Although the implemented scheduler employs a simple approach, it is able to shed some light on future research directions that consider the scheduling and precoding problem as a whole.

Finally in the simulations, the unrealistic assumption of a single GW multibeam HTS system is relaxed and the case where multiple GWs are serving the user beams is investigated. In this way, precoding is applied on the user beams served by each GW disregarding the interference coming from neighbouring beams served by the other gateways.

## System level simulation framework

The system simulator consists of two parts, namely a pre-processor and a simplified packet simulator. The pre-processor is responsible for generating all the physical layer related inputs needed by the packet simulator, like antenna gains and channel time series, as well as further inputs, like UT size. Instead, the simplified packet simulator is responsible for the clustering and scheduling of the users, the computation of the precoding matrix and the system level performance evaluation.

The simulator works on a time granularity based on PLFRAMES for the traffic generation and scheduling. During several PLFRAMES, the atmospheric effects are assumed to be constant. Traffic time series are generated for every user, indicating its active and idle periods. The traffic load per beam is a user defined parameter exploited for deriving the corresponding user load. In our simulations we assumed the beam traffic equally distributed among the users of each specific beam. The scheduler is responsible for the selection of the user to be served in the next PLFRAME and for the selection of the users that can be clustered together into one Bundled PLFRAME. The scheduling policy defines how the user is selected and this is done based on contrasting criteria and optimization targets, like overall system throughput or user fairness. For the simulations, a fair scheduling policy is chosen which aims at providing a similar throughput to users that experience different transmission channel conditions. In order to achieve fairness among the users, the first user of a frame is chosen based on the amount of unserved traffic. Since the users have an equal traffic request on average, users with a high amount of unserved traffic are likely to be in a bad channel condition, and thus are given priority over users that have less traffic waiting in the queues. After selecting the first user, the frame is filled with N-1 further users. These users are selected by the geographical user selection method [17] (see Section IV), which finds the users that have transmission channel conditions most similar to the first user.

The number of users that are encapsulated the same PLFRAME influences the precoding gain over a conventional system. According to [17], for computing the precoding matrix, the estimated channel states of the users in one frame are averaged. On the one hand, the more users are packed together into one frame, the more likely it is that the averaging introduces a not-negligible amount of estimation error. On the other hand, a higher number of users in one frame increase the frame utilization. This puts stronger requirements on the scheduler for finding a suitable trade-off for the cluster size. But this might also increase the scheduler complexity and thus the costs of the scheduling process. For keeping the scheduling complexity low in the simulations, the number of users that are selected for one frame is fixed and set to a not-optimized value of N = 5, which however is indicative of a typical IP packet size compared to the size of the bundled PLFRAME.

The system parameters chosen for the four colour benchmark scenario and the one colour precoding scenario are summarised in Table 1. The precoding scenario applies full frequency reuse with two polarizations per beam. In order to make fair comparisons, the saturated transmit power per beam is scaled appropriately. The feeder links between GWs and satellite are considered ideal in both cases.

Table 1: System level simulation parameters.

|  | Benchmark | Precoding |
|---:|---|---|
| **Colours** | 4 | 1 |
| **Downlink Frequency** | 20 GHz | 20 GHz |
| **Bandwidth** per Beam | 250 MHz | 500 MHz |
| **Number of Polarizations** | 2 per system | 2 per beam |
| **Number of Carriers** per beam | 1 | 1 |
| **Saturated RF power** per beam/polarization | 100 W | 50 W |
| **OBO** | 2 dB | 2 dB |
| **Roll-off** | 20% | 20% |
| **Terminal Antenna G/T** in clear sky | 16.9 dB/K | 16.9 dB/K |
| **ModCod** | DVB-S2x | DVB-S2x |
| **Users per beam** | 1000 | 1000 |

During the simulation, the actual served amount of traffic per user and the frame utilization are measured. The system capacity is computed using the served traffic and accumulated over all users in the system. The upper bound (u.b. in the figures) system capacity is computed assuming a full utilization of the frames, and provides an upper bound on the achievable throughput.

## System level simulation results

The system level simulations are carried out for two scenarios:

- a single GW scenario,
- a multi GW scenario.

In both cases, the system consists of 63 user beams covering central Europe mainly. In the single GW case, all 63 user beams are served by the same GW while in the multi GW case, nine GWs serve a cluster of seven beams each. For the latter case, the beam clustering, i.e. GW-user beams association is depicted in Figure 8. In each beam, 1000 users are uniformly distributed, resulting in an average user density of 0.023 users/km$^2$ inside the 3 dB coverage edge of every beam. The traffic request per beam ranges from 0.5 Gbit/s to 4.0 Gbit/s. This leads to a traffic rate per user between 1 Mbit/s and 8 Mbit/s on average (the peak data rate can be much higher), when the user is in the active state, as determined by the traffic time series.

The effect of channel estimation impairments are taken into account in the simulations and are compared to ideal channel estimation, when the channel matrix is assumed to be perfectly known at the receiver. On the one hand, the ideal channel estimation is able to estimate all signals independent from their strength and without any errors and provides always up-to-date values to the GW, which is responsible of computing the precoding matrix. On the other hand, a more realistic estimation is able to detect signals only up to a certain threshold, introduces errors on the estimated amplitude and phase, and the CSI may already be outdated when received by the GW, as explained in Section II.

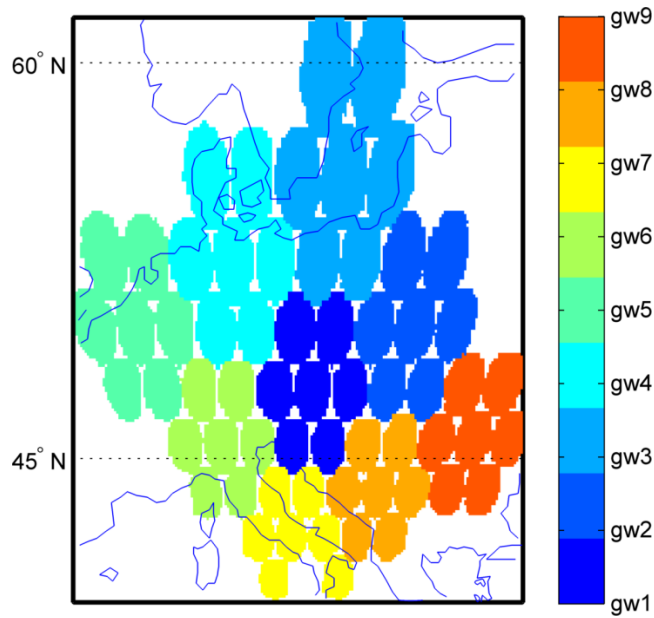

Figure 8: System simulations: Beam clusters per GW.

Three effects are taken into account in the simulations:

- Limit on the channel estimation for the interfering signals
- Effect of outdated phase estimation due to the round trip time (RTT)
- Effect of the estimation errors of both amplitude and phase on the channel matrix for the main signal as well as for the interfering signals

Limitations on the channel estimation as well as residuals from the estimation procedure for both amplitude and phase related to each waveform (the reference and the interfering waveforms) have been obtained through physical layer simulations. The physical layer impairments used in all simulations have been taken into account with respect to the DVB-S2x ESA channel model [23].

The effect of outdated phase estimation has been accounted starting from a realistic phase noise mask of a frequency converter operating in Ka-band. The residual standard deviation of the Gaussian approximation for the effect of outdated phase has been computed using the auto-correlation function assuming a RTT of 0.5 seconds. The exact numbers for the estimation errors and outdated phase are summarized in Table 2 for the two cases. A different estimation error is applied on the estimations for the main signal carrying the wanted information and for the interferer signals.

Table 2: Channel estimation impairments.

| | Ideal | Real |
|---|---|---|
| **Estimation threshold** log. | -Inf | -21 dB |
| **Outdated phase** std. deviation, lin. | 0° | 4.14° |
| **Main signal amplitude** mean / std. deviation; lin. | 0 / 0 | 0.0093 / 0.0143 |
| **Main signal phase** mean / std. deviation; lin. | 0 / 0 | -0.0115 / 0.0115 |
| **Interference amplitude** mean / std. deviation; lin. | 0 / 0 | 0.0064 / 0.0191 |
| **Interference phase** mean / std. deviation; lin. | 0 / 0 | 0.0102 / 0.0282 |

The simulation results for the single GW case are shown in Figure 9. The benchmark system is able to serve the requested traffic up to a rate of 55 Gbit/s. Up to this point, the offered traffic can be fully served. Beyond this point, the system starts to show saturation effects and not all of the offered traffic can be served resulting in packet losses. As the traffic rate grows, the capacity approaches the upper bound capacity of 71 Gbit/s. The precoding system with ideal CSI is able to serve the traffic up to a rate of 120 Gbit/s before saturation effects start to appear and packet losses occur. The upper bound capacity is around 170 Gbit/s. When introducing the effects of real CSI, the achievable outer bound capacity decreases to 120 Gbit/s. The region where all of the traffic can be served is also decreased to 70 Gbit/s.

The upper bound capacity for low traffic loads is higher than for high loads. This is due to the selected fair scheduling policy. A random scheduler, which selects the users randomly instead of according to the unserved traffic -not shown in the figure- would not show this slope. Instead, it would yield an almost constant system capacity that is even higher than the one achieved by the fair scheduler.

The fair scheduler selects the users according to their unserved traffic and thus prefers users with a worse channel than already served users with a better channel. When the traffic load is low, also the users with a worse channel can be completely served within one frame. As the load increases, these users have to be served in several frames, while for the users with better channel, a single frame may still be enough. This lowers the system capacity, but increases the fairness among users.

The implementation of a single GW system, especially with a high number of beams and users, and thus a high capacity demand, puts high demands on the feeder links. When switching to a more practical multi GW scenario with relaxed requirements on the feeder links, the achievable throughput is generally lower than in the single GW case, since only the signals transmitted by the GW's own beams can be precoded, but not the signals belonging to the other GWs' beams. The multi GW system capacity simulation results are depicted in Figure 10.

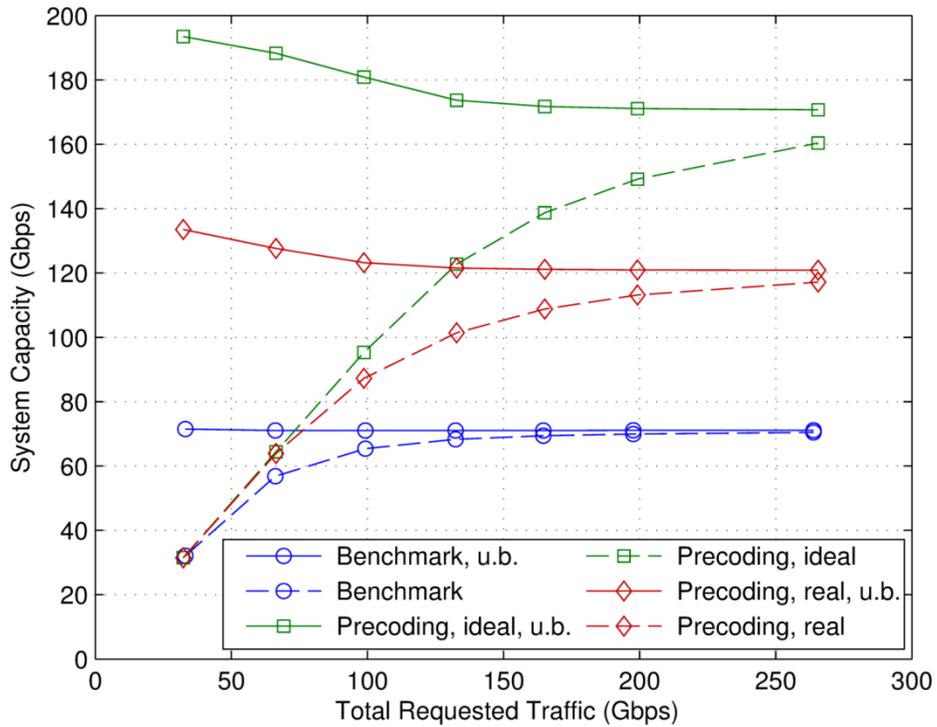

Figure 9: System capacity results, single GW.

The benchmark system in this case shows the same behaviour and achievable capacities like in the single GW case. This is expected, since the system is manly interference limited and the interference figures do not change from the single to the multi-GW scenario. The two precoding systems start saturating at around 60 Gbit/s, which is around the same point as the benchmark. But as the traffic load increases, the precoding system capacity keeps increaseing, although at a smaller slope. The upper bound capacity for the precoding scheme with ideal CSI is 113 Gbit/s, while the realistic CSI model causes a performance reduction down to 97 Gbit/s. Anyway, the precoding capacity still outperforms the one achieved by the benchmark system. The effect introduced by the scheduling policy, a slightly increased upper bound capacity, can also be observed in the multi GW case, exactly like in the single GW case.

The system capacity gain achieved by precoding relative to the benchmark system is illustrated in Figure 11 for all previously presented simulation scenarios. They are further summarized in Table 3. In the presence of a single GW with ideal CSI, the upper bound capacity is 1.4 times higher than in the benchmark system, and even with the realistic CSI model, the capacity can be increased by 70% compared to the four colour benchmark system. In the multi GW case, the out-of-cluster interference coming from the rest of the beams leads to capacity gains of 58% for ideal CSI and 38% with real CSI estimation models. It is worth repeating that these results refer to multicast precoding with 5 users encapsulated in each frame.

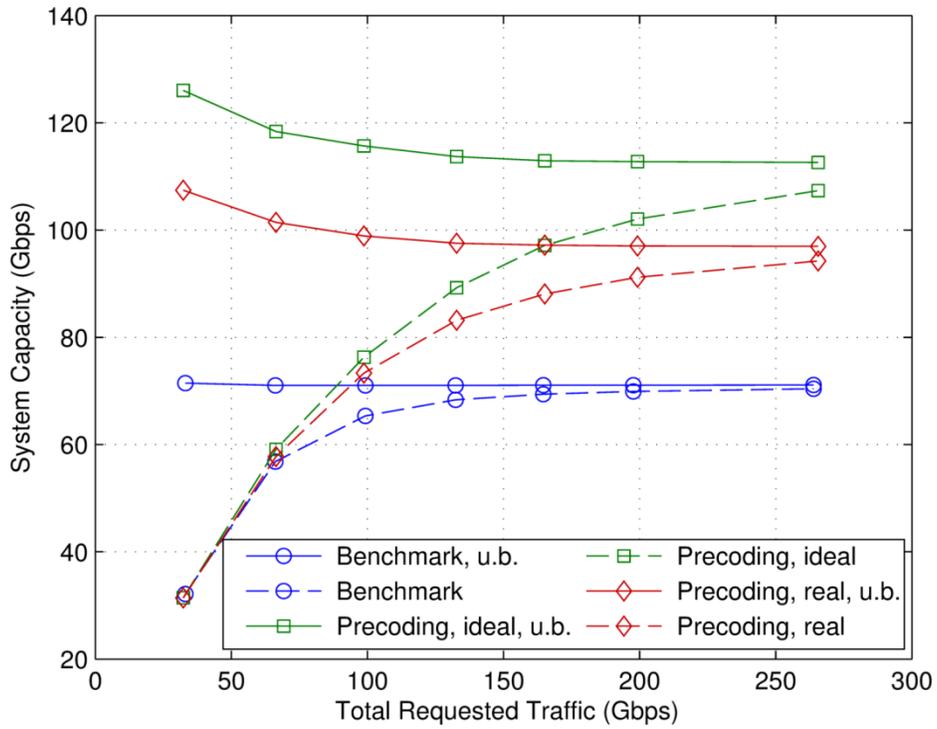

Figure 10: System capacity results, multi GW case.

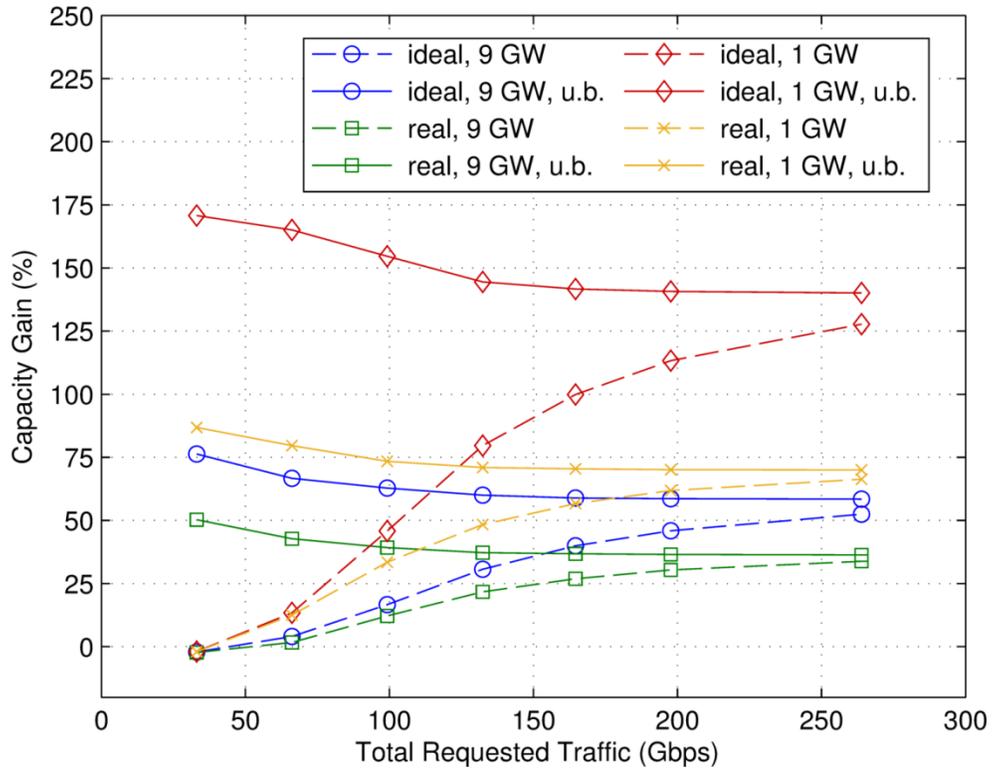

Figure 11: Precoding capacity gain.

Table 3: Performance summary of multicast precoding with 5 users per frame.

|  | Upper bound Capacity (Gbit/s) | | Gain |
| --- | --- | --- | --- |
|  | Benchmark | Precoding |  |
| Single GW, ideal | 71.17 | 170.69 | 140% |
| Single GW, real | 70.85 | 120.40 | 70% |
| Multi GW, ideal | 71.17 | 112.62 | 58% |
| Multi GW, real | 70.85 | 97.48 | 38% |

## VII. Conclusions

Similar to the MU-MIMO approach in LTE, DVB-S2x has laid the groundwork for advanced interference management techniques to be applied in multibeam HTS systems allowing for a quantum leap in terms of offered system capacity. This paper discusses the main practical issues that are encountered when considering the application of precoding in high frequency reuse multibeam HTS systems. Most of these issues are tackled by the new superframe specification of DVB-S2x, which is briefly overviewed in the context of the precoding technique. Furthermore, the paper summarizes the advanced synchronization algorithms that have been developed in recent ESA funded R&D to overcome the challenges of imperfect and outdated channel estimation. The realistic system simulator developed for this purpose, including the dimension of traffic generation and user scheduling, testifies to precoding gains that range from 40% up to 140% depending on the system scenario, even after accounting for the practical issues identified throughout the paper. These gains are achieved by applying simple user scheduling and multicast precoding approaches for 5 users per frame. Further gains are to be expected through optimizing both these two operations.

Further work is required to compare the performance of precoded networks with classical frequency re-use 4 systems at a given maximum power and payload mass. Possibly, this might lead to precoded HTS networks using less conventional payload architectures with semi-active antennas and SSPA technologies.

## List of Acronyms

ACM:    Adaptive Coding and Modulation

APSK:   Amplitude Phase Shift Keying

BCH:    Bose Chaudhuri Hocquenghem

CDI:    Channel Direction Information

CSI:    Channel State Information

CQI:    Channel Quality Information

| | |
|---|---|
| DTH: | Direct-to-Home |
| DVB-S2(x): | Digital Video Broadcasting via Satellite Second Generation (Extension) |
| ESA: | European Space Agency |
| FSS: | Fixed Satellite Service |
| GEO: | Geostationary |
| GW: | Gateway |
| HPA: | High Power Amplifier |
| HTS: | High Throughput Satellite |
| LDPC: | Low Density Parity Check |
| LNB: | Low Noise Block |
| LO: | Local Oscillators |
| LTE: | Long Term Evolution |
| MMSE: | Minimum Mean Square Error |
| ModCod: | Modulation & Coding |
| MU-MIMO: | Multi User - Multiple Input Multiple Output |
| NGW: | Next Generation Waveform |
| PLFRAME: | Physical Layer Frame |
| PLH: | Physical Layer Header |
| RTT: | Round Trip Time |
| SNIR: | Signal-to-Noise and Interference Ratio |
| SoSF: | Start-of-Superframe |
| SSPA: | Solid State Power Amplifiers |
| SU-MIMO: | Single User – Multiple Input Multiple Output |
| TM: | Transmission Mode |
| UT: | User Terminal |